\documentclass[ preprint,
                superscriptaddress, 
                aps, 
                prx, 
                noeprint,
                citeautoscript]{revtex4-2}
\usepackage{graphicx}
\usepackage{dcolumn}
\usepackage{bm}
\usepackage{amsmath,amssymb}
\usepackage[version=4]{mhchem}
\usepackage{xcolor}
\usepackage{siunitx}
\DeclareSIUnit\angstrom{\text {Å}}
\usepackage{braket}
\usepackage{xr}
\usepackage{float}
\usepackage{multirow}
\usepackage{wrapfig}
\usepackage{gensymb}
\usepackage{quotes}
\usepackage{xr}
\usepackage{gensymb}

\begin{document}

\title{Temperature-Induced Crossover of Coherent Phonon Mechanisms in Chiral 2D Perovskites}

\author{Katherine~A~Koch}
\email{kochka22@wfu.edu}
\affiliation{Department of Physics \& Center for Functional Materials, Wake Forest University, Winston-Salem, NC~27109, United~States}

\author{Matthew~P~Hautzinger}
\affiliation{National Laboratory of the Rockies, Golden, Colorado 80401, United States}

\author{Matthew~C~Beard}
\affiliation{National Laboratory of the Rockies, Golden, Colorado 80401, United States}

\author{Ajay~Ram~Srimath~Kandada}
\email{srimatar@wfu.edu}
\affiliation{Department of Physics \& Center for Functional Materials, Wake Forest University, Winston-Salem, NC~27109, United~States}

\date{\today}

\begin{abstract}
The coupling between electronic excitations and lattice degrees of freedom fundamentally dictates the optoelectronic functionality of hybrid perovskites. While the potential energy surfaces (PESs) of the electronic excited states are typically considered static, albeit modulated by thermal disorder, the exact nature of their structural evolution with temperature remains elusive. Here, we demonstrate that the excited-state structural reconfiguration in two-dimensional metal-halide perovskites is explicitly temperature-evolving, governed by lattice compliance. We select a chiral perovskite framework with an exceptionally large, temperature-dependent bond angle variance to maximize the structural compliance. Through phase-resolved resonant impulsive stimulated Raman scattering, we measure the coherent phonon dynamics and resolve the real-time structural pathways of exciton-lattice dressing. We observe a temperature-induced crossover from field-driven Impulsive Stimulated Raman Scattering (ISRS) to population-driven Displacive Excitation of Coherent Phonons (DECP). While momentum-driven ISRS pathways dominate at low temperatures, increasing thermal energy softens the lattice and enhances coordinate-driven displacive pathways, allowing excitons to sample steeper, highly anharmonic regions of the excited-state PES. Our results show that temperature can actively modulate the excited-state structural coordinates of flexible 2D frameworks, offering a practical strategy to tune exciton-lattice interactions in chiral optoelectronics.

\end{abstract}

\maketitle

\section{Introduction}
The photophysical properties of two-dimensional metal-halide perovskites (2D-MHPs) is fundamentally determined by the structural and dynamic fluctuations of their inorganic frameworks. In these quantum confined systems, strong excitonic resonances are profoundly dressed by lattice distortions, resulting in pronounced polaronic effects that dictate carrier lifetimes, effective masses, and radiative behaviors~\cite{thouin2019phonon, srimath2020exciton, biswas2024exciton}. Conventionally, the structural landscape governing these interactions has been treated as static at low temperatures~\cite{zeiske2022static, baranowski2018static}, with rising thermal energy introducing stochastic dynamic disorder that broadens electronic transitions and accelerates phase decoherence~\cite{munson2021influence, munson2018dynamic, srimath2022homogeneous, duan20242d}. Nevertheless, the fundamental topology of the underlying potential energy surfaces (PESs) is rarely static; rather, the lattice's configurational space can evolve significantly with temperature, dynamically altering the energy gradients that dictate light-matter interactions. Directly observing such temperature-evolving PES behavior requires a structural platform possessing an exceptionally flexible, highly distorted framework, along with a spectroscopic probe capable of tracking how specific structural relaxation pathways scale with thermal energy.

Chiral 2D-MHPs offer a unique structural platform to address this challenge. While heavily investigated for their promise in emerging chiroptoelectronics and spin-polarized transport~\cite{billing2006synthesis, ahn2017new, dong2025chirality, ma2021chiral, li2024enhancing, lu2020highly, hautzinger2024room}, the profound structural consequences of introducing bulky, symmetry-breaking chiral spacer cations are of paramount thermodynamic interest~\cite{haque2025remote, li2024large, valev2013chirality, ma2021recent, jana2020organic,lu2020highly, ahn2017new, ma2021chiral, ma2022elucidating, ding2024structure}. The asymmetric packing constraints of these chiral organics force severe octahedral distortions within the inorganic network. Crucially, the resulting bond angle variance (BAV) in systems such as chiral (R/S-) methylbenzylammonium lead iodide, \ce{(R/S-MBA)2PbI4} is anomalously large ($\sigma^2 \sim 20$)—significantly exceeding that of prototypical, achiral analogues like phenethylammonium lead iodide, \ce{(PEA)2PbI4} ($\sigma^2 \sim 3$). This deliberate amplification in the static structural distortion generates a highly compliant and anharmonic inorganic lattice, where the equilibrium framework is extraordinarily sensitive to thermal fluctuations. Most importantly, this exceptional BAV is not invariant but increases progressively with increasing temperature, making chiral perovskites an ideal testing ground for tracking the thermal evolution of potential energy landscapes.

To map these changing landscapes, tracking real-time coherent structural responses offers a unique solution. Impulsive optical excitation initiates coherent lattice vibrations via two pathways: Impulsive Stimulated Raman Scattering (ISRS), a field-driven process that follows the ground-state configurations, and Displacive Excitation of Coherent Phonons (DECP), a population-driven mechanism sensitive to excited-state gradients~\cite{dhar1994time, kumar2001investigations}. Because ISRS scales with electronic phase coherence while DECP directly reflects nuclear structural displacements on the excited-state potential energy surface (PES), the dynamic competition between these two pathways serves as a sensitive probe of the evolving PES architecture.

In this work, we employ phase-resolved resonant impulsive stimulated Raman scattering (RISRS)~\cite{thouin2019phonon} across a wide temperature range to demonstrate a temperature-induced crossover in these coherent phonon mechanisms. We demonstrate that as temperature rises, the severe, thermally expanding BAV of the \ce{(R/S-MBA)2PbI4} lattice reshapes the excited state lattice configuration. While the momentum-driven ISRS pathways decay rapidly due to the thermal erosion of electronic phase coherence, the displacive DECP pathways exhibit a relative enhancement. This behavior reveals that rising thermal energy acts as a structural filter. By softening the lattice and expanding the BAV, it allows the photogenerated excitons to sample steeper, highly anharmonic regions of the excited-state PES. Ultimately, these findings indicate that the potential energy surfaces in hybrid frameworks are explicitly temperature-evolving, in which chiral-induced structural flexibility determines the precise vectorial directions along which the lattice reorganizes in response to light. 

\section{Results}

\subsection{Linear and Transient Absorption}

\begin{figure}[h]
    \centering
    \includegraphics[width=\columnwidth]{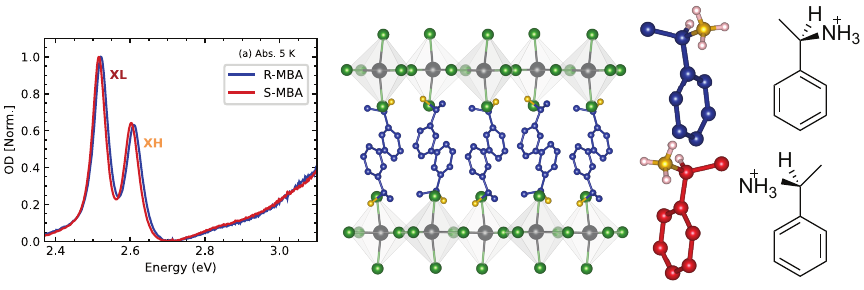}
    \caption{(a) Linear absorption of \ce{(R/S-MBA)2PbI4} measured at 5\,K. (Right) Structure schematic of the material system and the chemical structure of the two chiral enantiomers.}
    \label{fig:Abs_Structure}
\end{figure}

The low-temperature (5\,K) linear absorption spectrum of \ce{(R/S-MBA)2PbI4}, Figure~\ref{fig:Abs_Structure}(a), reveals a well-resolved excitonic structure characterized by two distinct resonances (XL $\approx$2.52\,eV and XH $\approx$ 2.61\,eV). Intricate fine structures at low temperature are typical for 2D-MHPs, sparking significant debate regarding their origin. Similar features in \ce{(PEA)2PbI4} have been attributed to phonon replicas~\cite{straus2016direct, dyksik2024polaron}; however, the energy separation observed here ($\approx$ 90 meV) is much larger than the energy separation reported for PEA-based systems ($\approx$ 35 meV). Hence, it is unlikely to be a simple vibrational sideband of XL and instead suggests a more fundamental electronic or structural origin. 

The dual-resonance structure is observed in all three sample variations \ce{racemic/R/S-MBA}. Figure~\ref{fig:Abs_Structure}(a) confirms that the spectral positions, line shapes, and relative intensities of XL and XH remain identical, barring sample-to-sample variation, across the two chiral enantiomers, suggesting the behavior is intrinsic to the MBA-based Pb-I framework, ruling out defect-related scenarios. The X-ray diffraction pattern shows no evidence of multiple crystalline configurations or phases, as only extremely minor phase impurities are detected (Fig.~\ref{supp-fig:XRD}). Another possible scenario is that the peaks represent two unique excitonic states within the same crystal structure, possibly originating from different conduction band valleys or distinct polaronic couplings, as has been suggested in alternative perspectives for other 2D-MHPs~\cite{srimath2020exciton, thouin2019phonon, koch2025structure}.

\begin{figure}
    \centering
    \includegraphics[width=10cm]{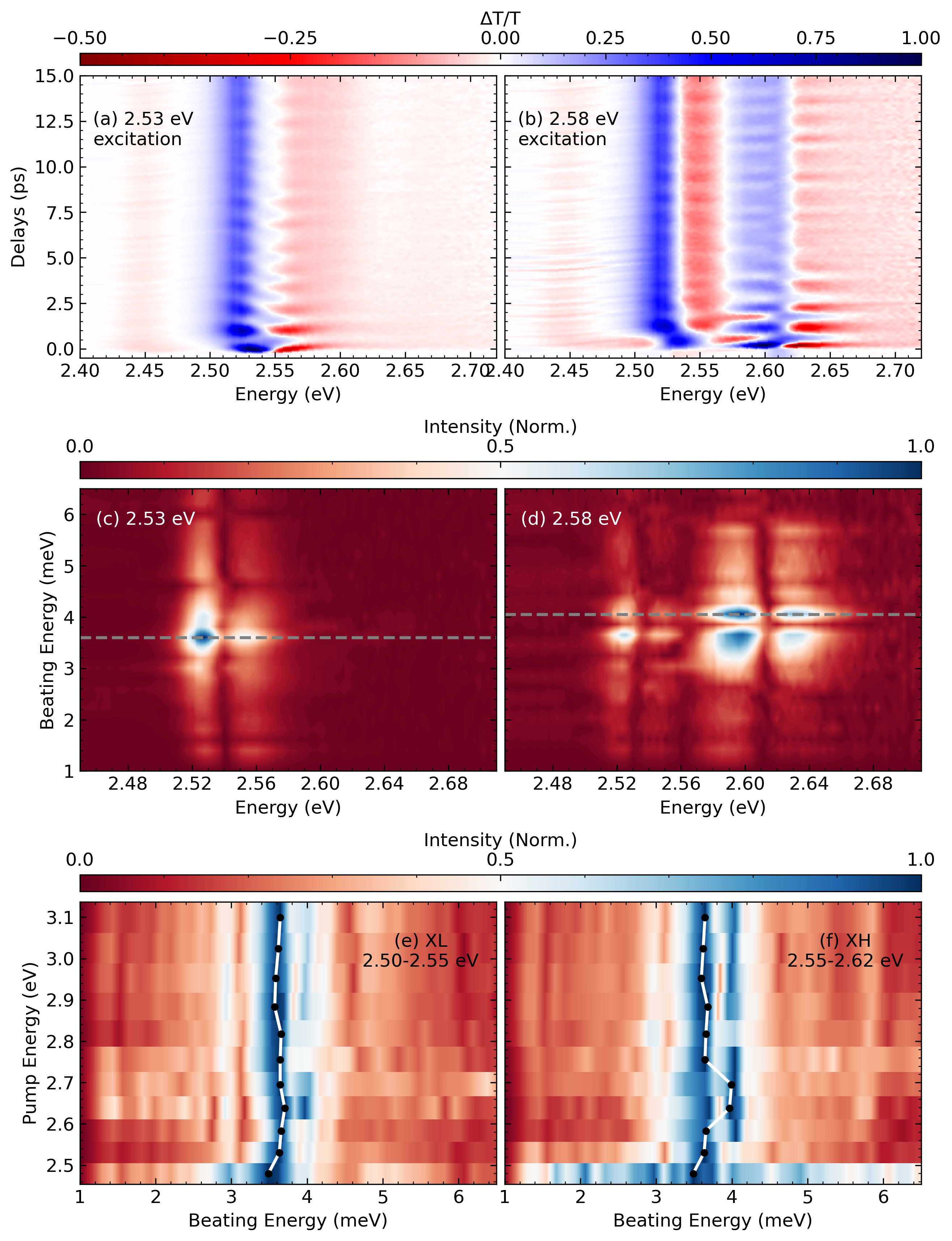}
    \caption{Transient absorption (TA) spectra of \ce{(R-MBA)2PbI4} measured at 15\,K, with an approximate fluence of $\sim$17$\mu$J/cm$^2$ with a pump energy of (a) 2.53\,eV and (b) 2.58\,eV, resonant with the XL and XH transition, respectively. The corresponding beating spectra are shown in panels (c) and (d), respectively. Probe integrated Raman spectra for (e) XL and (f) XH as a function of pump energy.}
    \label{fig:TA_Raman}
\end{figure}

To further investigate the relationship between the XL and XH resonances, we analyze a series of pump-energy dependent TA measurements at 15\,K. By selectively photo-exciting the system at energies resonant with both the lower (XL) and higher (XH) manifolds, we analyze the resulting spectral lineshapes and carrier dynamics to determine if these states are electronically coupled. 
Notably, the resulting TA maps, shown in Fig.~\ref{fig:TA_Raman}(a) and (b), reveal markedly different spectral responses depending on the excitation conditions. Resonantly pumping the XL transition (2.53\,eV - Fig.~\ref{fig:TA_Raman}(a)) reveals a relatively selective response, dominated by a ground-state bleach (GSB) at the XL energy. Superimposed on this bleach is a derivative-like lineshape suggesting a photo-induced shift of the XL resonance under excitation. In addition, a red-shifted photoinduced absorption (PA) band emerges below the XL energy, indicative of an excited state absorption (ESA)~\cite{berera2009ultrafast}. Spectral signatures with the XH resonance are absent, implying that excitation at 2.53\,eV does not access higher-energy states, and XL and XH may not share a common ground state or continuum of intermediate states~\cite{jia2024transient}. In contrast, excitation at 2.58\,eV, resonant with the XH transition, produces a richer TA response (Fig.~\ref{fig:TA_Raman}(b)). Multiple features emerge, including a similar low-energy ESA feature, two distinct GSB signals (XL and XH), and several PA bands. The presence of both XL and XH bleaches under this excitation condition suggests that 2.58\,eV populates a broader manifold of states, enabling access to multiple excitonic transitions. 
Spectral decomposition, see Section~\ref{supp-sec:TA_lineshape} of the SI for more details, reveals that XL excitons dominate the photoresponse across a broad range of pump energies, while XH excitons are only accessed under resonant excitation. Additionally, unique photoinduced energy shifts are observed for the two transitions (blue shift for XL and red shift for XH), pointing to fundamentally different mechanisms governing their behavior, possibly originating from transitions in different regions of the carrier reciprocal space, such as distinct conduction-band minima or valleys.

\subsection{Resonant Impulsive Stimulated Scattering}

To understand the impact of lattice dynamics on these resonances, we employ resonant impulsive stimulated Raman scattering (RISRS), which allows for the real-time observation of coherent lattice vibrations and their dynamic interaction with electronic states~\cite{dhar1994time, merlin1997generating, thouin2019phonon}. In this experiment, an optical pulse with a duration shorter than the period of the vibrational mode ($t_{pulse} < 1/\nu$) \textit{impulsively} drives the system, creating a coherent superposition of vibrational states, manifesting as a nuclear wavepacket. These coherent nuclear oscillations periodically modulate the system's electronic transition energies and oscillator strengths. Isolation of the vibrational dynamics and a Fourier transform along the delay axis, see Section~\ref{supp-sec:RISRS_Data}, yields a two-dimensional beating map correlating the electronic and vibrational landscapes (Figs.~\ref{fig:TA_Raman}(c)\&(d)).

Much like the TA response, the distribution of vibrational coherence across the excitonic landscape is highly sensitive to the excitation energy (Fig.~\ref{fig:TA_Raman}). Exciting the XL transition confines modulation to the XL probe energy; conversely, a pump resonant with XH triggers modulation across both regions, though the XH response is more prominent. This selective enhancement demonstrates that the coherent phonon signatures in the transient response are determined by the specific electronic state populated by the impulsive excitation, thereby providing a clear map of state-specific exciton–phonon coupling. Furthermore, the presence of two distinct modulation nodes and a unique dominant phonon mode energy associated with XL and XH (Fig.~\ref{fig:TA_Raman}(c)\&(d)) can be interpreted within the exciton-polaron framework. This suggests that XL and XH correspond to separate excitonic species that couple to the lattice along unique coordinates, with the branching ratio of these coupling pathways being strictly dictated by the initial impulsive excitation energy~\cite{thouin2019phonon, biswas2024exciton, koch2025structure}. Additionally, the 2D-maps presented in Fig.~\ref{fig:TA_Raman}(e)\&(f) display the pump energy-dependent Raman spectra, highlighting the complexities of the potential energy surfaces in the electronic excited states of \ce{(R-MBA)2PbI4}.

\begin{table}[h]
    \centering
    \begin{tabular}{c|c|c|c|c}
        $M1$ & $M2$ & $M3$ & $M4$ & $M5$ \\
        \hline  1.9\,meV & 2.5-2.7\,meV & 3.0-3.3\,meV & 3.5-3.9\,meV & 4.1-4.5\,meV
    \end{tabular}
    \caption{Dominant phonon mode energies for \ce{(R-MBA)2PbI4}}
    \label{tab:modes}
\end{table}

The spectral profile of the coherent oscillations, referred to as the \textit{modulation lineshape}, provides a direct probe of how specific lattice vibrations are coupled to the excitonic transitions. This lineshape, as seen in Fig.~\ref{fig:Raman_phase}(c)-(g), is obtained by estimating the amplitude (blue lines) and the relative phase (orange/dark red circles) of the oscillations at each of the phonon frequencies across the probe spectrum. For systems with strong exciton–phonon interaction, the modulation spectrum exhibits a pronounced dip at the excitonic resonance and a corresponding $\pi$ phase jump~\cite{thouin2019phonon}. The depth, width, and the overall lineshape of the modulation are governed by the strength of the lattice displacement induced by the excitonic excitation~\cite{pollard1992analysis}. In particular, this lineshape can be related to the Huang-Rhys parameter, $\mathbf{S}=\frac{1}{2}\Delta^2$, quantifying the lattice displacement ($\Delta$) in the excited state potential energy surface (PES) induced by exciton-phonon coupling~\cite{barclay2022characterizing, turner2020basis, arpin2021signatures} (Fig.~\ref{fig:Raman_phase}(b)). The corresponding phase profile of the phonon mode yields information about the excitation mechanism and nature of the observed coherent phonon.  

\begin{figure}[h]
    \centering
    \includegraphics[width=\columnwidth]{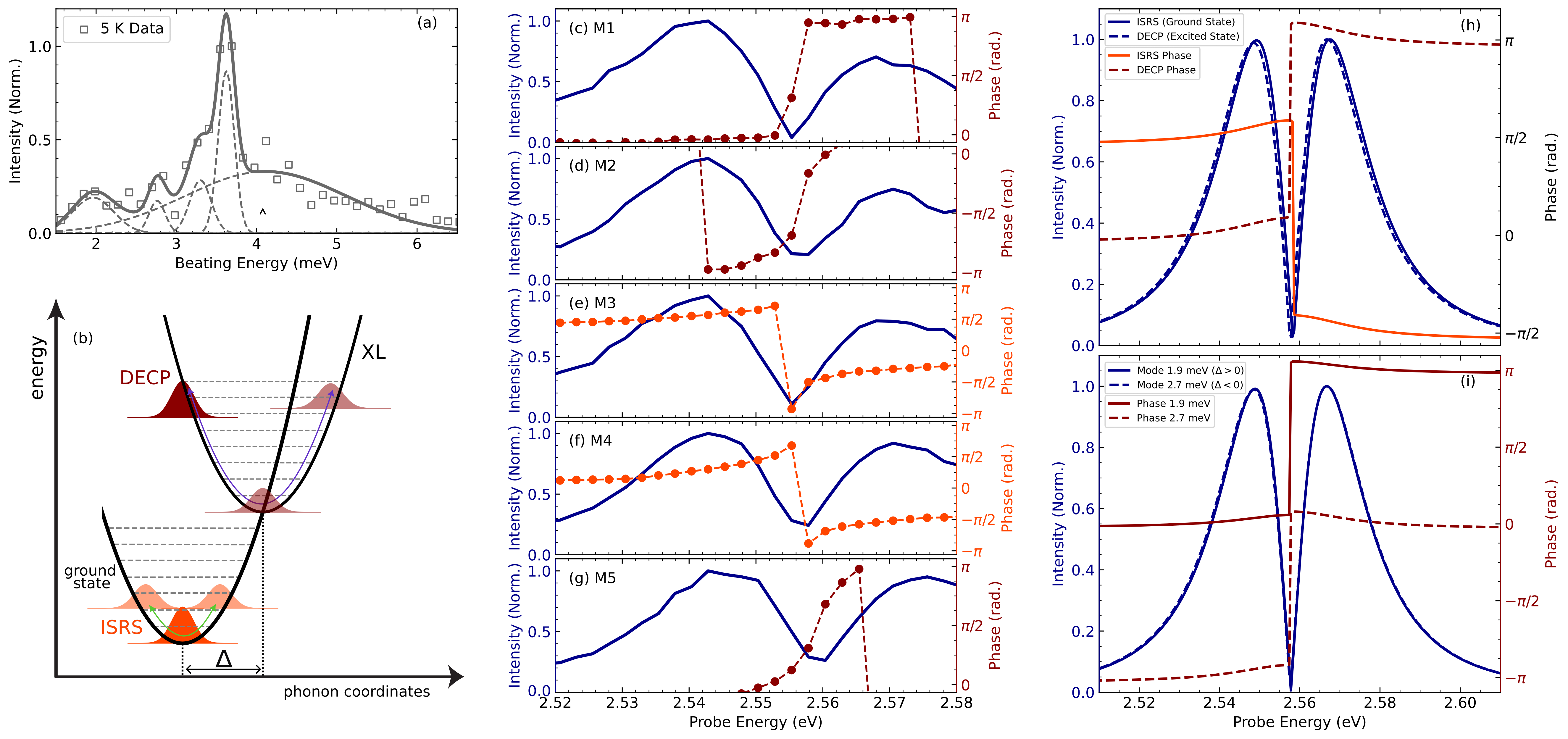}
    \caption{(a) The Raman spectrum of \ce{(R-MBA)2PbI4} measured at 5\,K with a pump energy of 2.53\,eV (squares), resonant with the XL transition, and the corresponding Gaussian fit (solid lines with components shown with dashed lines). (b) Schematic illustration of the two possible phonon excitation pathways, impulsive stimulated Raman Scattering (ISRS) and displacive excitation of coherent phonons (DECP). (c)-(g) The modulation spectra (dark blue line) and corresponding phase profiles (orange (dark red) circles for ISRS (DECP) modes) of the various phonon modes observed in the beating map (Fig.~\ref{supp-fig:SI_Raman_Phase}) and corresponding Raman spectrum in panel (a). Simulated modulation spectra and phase profiles utilizing the model outlined in Ref.~\citenum{kumar2001investigations} comparing (h) ISRS and DECP modes and (i) DECP modes with positive and negative lattice displacements ($\Delta$), where $\Delta$ is related to the Huang-Rhys parameter ($S=\Delta^2/2$).} 
    \label{fig:Raman_phase}
\end{figure}

Impulsive excitation of coherent phonons is primarily described by two models (see schematic in Fig.~\ref{fig:Raman_phase}(b)): impulsive stimulated Raman scattering (ISRS) and displacive excitation of coherent phonons (DECP). In the framework of ISRS, the pump pulse acts as a resonant field that creates a ground-state coherence ($\delta\rho_{gg}$) through resonant coupling with the excited state manifold. This process imparts an impulsive momentum kick ($P_0 \neq 0$) to the nuclei while they remain at their equilibrium coordinates ($Q_0 \approx 0$), resulting in a wavepacket that initiates its motion with maximum velocity and a sinusoidal temporal phase. Conversely, DECP creates a real excited-state population ($\delta\rho_{ee}$), where the change in the electronic distribution abruptly shifts the potential energy minimum of the lattice~\cite{yan1991pulse, yan1989ultrafast, mukamel1990femtosecond, yan1990femtosecond, kumar2001investigations}. Because the nuclei are dropped at the Frank-Condon point, far from the new excited-state equilibrium, the wavepacket is initiated with a finite displacement ($Q_0 = \Delta$) and zero initial momentum ($P_0 \approx 0$), resulting in a cosine-like temporal response~\cite{nakamura2015influence}. The $\pi/2$ phase difference observed in the modulation phase profile is a direct manifestation of this fundamental distinction between momentum-driven (ground-state) and position-driven (excited-state) coherence.

As illustrated in Figs.~\ref{fig:Raman_phase}(c)-(g), every observed mode exhibits the characteristic $\pi$ phase jump across the modulation dip. Superimposed on this modulation amplitude profile, which remains the same for all the observed modes, a clear distinction emerges in the relative phase profiles. The two dominant modes, $M3$ and $M4$ (Fig.~\ref{fig:Raman_phase}(e)\&(f)) display a phase decrease across the modulation dip that aligns with the absorption peak energy of the XL transition. The remaining modes (Fig.~\ref{fig:Raman_phase}(c),(d)\&(g)), however, exhibit a relative phase offset of approximately $\pi/2$ with that of the primary Raman modes. 

To rationalize these trends, we simulated the effective linear response of the vibrational coherences following the theoretical framework of Kumar et al.~\cite{kumar2001investigations} (details provided in Section~\ref{supp-sec:Kumar_simulations} of the SI). Our simulations confirm that this $\pi/2$ offset is the \textit{smoking gun} for distinct initial conditions in Liouville space. $M3$ and $M4$ are driven by an impulsive momentum kick ($P_0 \neq 0, Q_0 \approx 0$) on the electronic ground-state potential (ISRS-driven). In contrast, the remaining modes are initiated by an abrupt coordinate shift ($Q_0 = \Delta, P_0 \approx 0$) upon electronic excitation, consistent with a DECP mechanism (Fig.~\ref{fig:Raman_phase}(h)). This phase-based classification provides a robust method to disentangle ground- and excited-state dynamics that are otherwise spectrally overlapping.

The DECP group further exhibits an internal phase dichotomy, a $\pi$ shift between $M1$ and $M2$. Within the displaced harmonic oscillator framework, this phase inversion between DECP modes can be attributed to the direction of the potential displacement. By assigning a positive Huang-Rhys parameter ($\Delta > 0$) to $M1$ and a negative parameter ($\Delta < 0$) to $M2$, the simulation perfectly reproduces the experimental phase difference (Fig.~\ref{fig:Raman_phase}(i)). This suggests that the electronic transition triggers a stereospecific structural relaxation, where the lattice simultaneously expands along the $M1$ coordinate while contracting along the $M2$ coordinate, or vice-versa. Consequently, the phase profiles provide a direct vectorial map of the structural dynamics, identifying both the excitation pathway and the specific direction of lattice reorganization.

The disparity in the spectral linewidths between the two groups -- DECP modes are consistently broader than ISRS modes -- reflects a fundamental difference in their respective dephasing environments. In the ground-state manifold (ISRS), phonon lifetimes are primarily limited by phonon-phonon scattering, governed by the structural anharmonicity of the lattice at equilibrium. In contrast, DECP modes evolve on the excited-state PES and are subject to an additional and often dominant dephasing channel: exciton-phonon scattering~\cite{thouin2019enhanced, mauck2019excitons, rojas2023many, rojas2022peculiar}. Because DECP is a population-driven process, the coherent nuclear wavepacket coexists with a dense population of photogenerated excitons. This creates a high-probability scattering environment where the vibrational coherence is destroyed by stochastic fluctuations from electronic excitations. The increased linewidth thus serves as a spectroscopic signature of the enhanced anharmonicity and shortened coherence times inherent to the excited-state manifold, where the lattice is subject to phonon and electronic many-body interactions.

\begin{figure}[h]
    \centering
    \includegraphics[width=13cm]{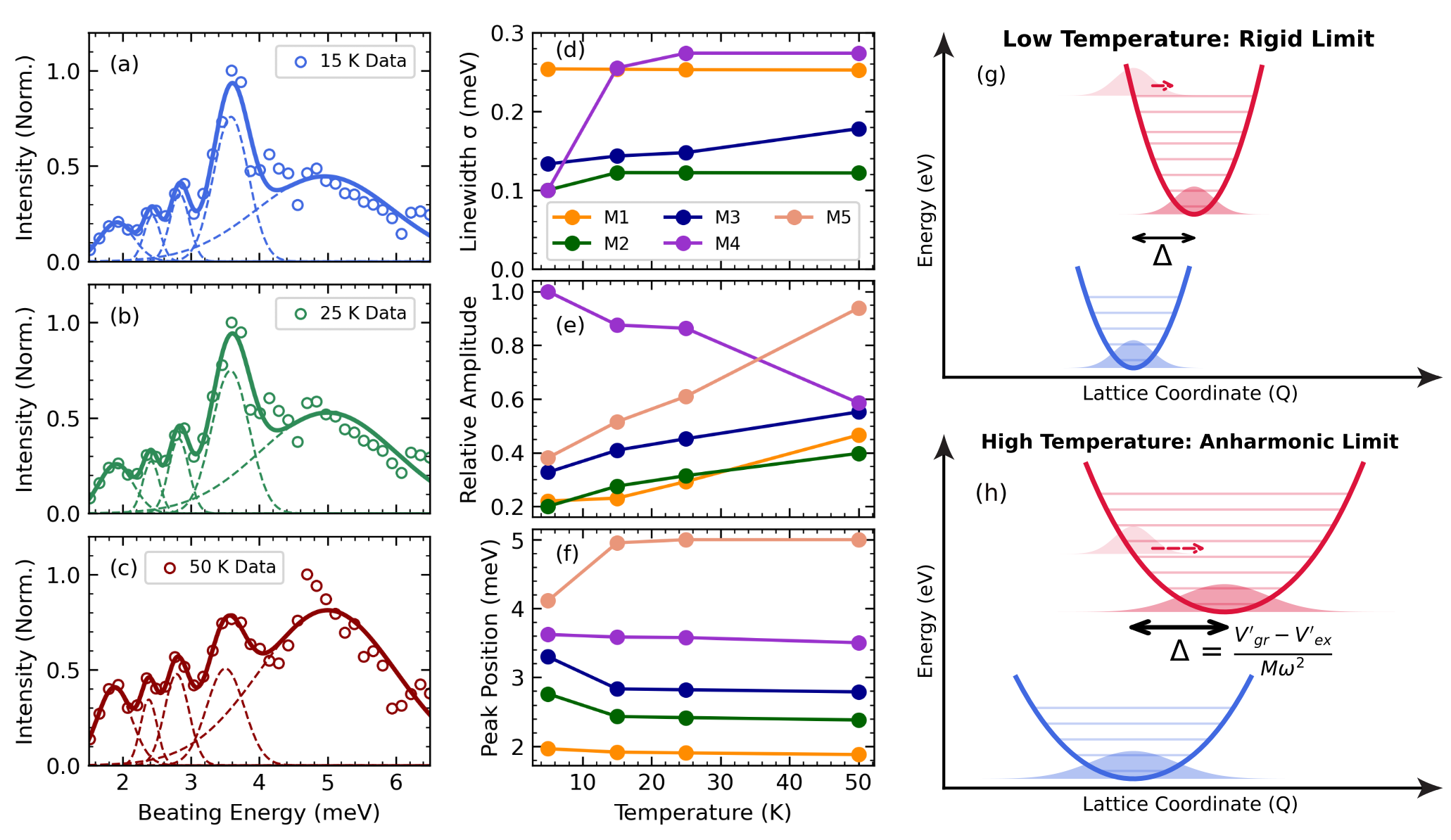}
    \caption{Probe-energy-integrated vibrational spectra (circles) for \ce{(R-MBA)2PbI4} measured at (a) 15\,K, (b) 25\,K, and (c) 50\,K, and their corresponding fits (solid line) where the underlying fitting components are represented with dotted lines. (d) Linewidth evolution of the various Gaussian peaks used in the fitting as a function of temperature. (e) The relative amplitudes of the various Gaussian peaks used in the fitting as a function of temperature. (f) Peak position of the various Gaussian peaks used in the fitting as a function of temperature. Potential energy surface (PES) schematic in the (g) low and (h) high temperature limit. The lattice is more rigid in the low temperature configuration, while in the high temperature limit, the system samples a softer, more anharmonic PES, corresponding to a relative increase in the DECP mode intensities.}
    \label{fig:Raman_temp}
\end{figure}

The modulation spectral profiles (Fig.~\ref{supp-fig:SI_Mod_Temp_Dep}) and corresponding Raman spectra (Fig.~\ref{fig:Raman_temp}(a)–\ (c)) exhibit a notable temperature dependence, where the amplitude of the modulation evolves with increasing temperature. The higher energy tail of the modulation diminishes, suggesting a temperature-dependent modification of exciton–lattice coupling and, consequently, of the effective Huang–Rhys factor. This apparent evolution initially seems inconsistent with the Born Oppenheimer approximation, as the Huang-Rhys parameter and deformation potential are governed by the electronic structure and bonding geometry, which are typically temperature-independent in the absence of lattice phase transitions~\cite{zhang2019applications, mukamel1995principles}. 

However, in 2D HOIPs, temperature-dependent lattice fluctuations driven by dynamic disorder introduce a layer of complexity that effectively renormalizes these coupling parameters. Dynamic disorder arises from thermally activated fluctuations in the orientations of the organic cations, leading to a temporally evolving distribution of local lattice geometries~\cite{yaffe2017local, munson2021influence, bakulin2015real, yang2026lattice}. In this regime, the deformation potential and the associated Huang–Rhys factor are interpreted as effective, ensemble-averaged quantities that reflect the range of accessible configurations sampled by the lattice. Structural distortion, intrinsic to the equilibrium structure and quantified through the bond angle variance ($\sigma^2$), which measures deviations from an ideal 90\degree \ bond angle, must also be considered. While BAV measures a uniform, periodic deviation from ideal octahedral symmetry rather than spatial static disorder, such severe ground-state distortion acts as a direct precursor to enhanced lattice compliance. For R/S-MBA, $\sigma^2$ is substantially larger ($\sim$ 20, see Fig.~\ref{supp-fig:crystal_params}) than that of a prototypical 2D-MHP, \ce{(PEA)2PbI4} ($\sim$ 3), indicating a significantly more distorted inorganic framework~\cite{koch2025structure}. Crucially, despite the anomalously large BAV ($\sigma^2 \sim 20$), the observation of sharp, well-resolved excitonic transitions ($XL$ and $XH$) at low temperatures confirms that this metric reflects a coherent structural distortion rather than random static disorder, which would otherwise broaden or wash out these fine features. Instead, this static distortion sets the stage for a highly responsive potential energy landscape that undergoes thermal renormalization as lattice flexibility scales with temperature.

The increase in the excited-state displacement with increasing temperature can be derived by examining the temperature dependence of the relative amplitudes of the ISRS and DECP features identified earlier. As shown in Fig.~\ref{fig:Raman_temp}(d), the linewidths of the Raman modes, except for the dominant ISRS mode, remain nearly constant with temperature, indicating no significant increase in homogeneous or inhomogeneous broadening~\cite{toyozawa2003optical, mauck2019excitons}. Interestingly, the relative amplitudes of the Raman modes exhibit a divergent temperature dependence (Fig.~\ref{fig:Raman_temp}(e)). While the absolute modulation amplitudes decrease across the board due to increased thermal decoherence, the DECP-driven modes gain relative intensity compared to the dominant ISRS mode as temperature increases. This distinct temperature-dependent behavior of the ISRS and DECP modes reveals a fundamental divergence in how ground- and excited-state coherences interact with the evolving lattice landscape, which can be rationalized through the initial wavepacket moments.

In the impulsive limit, the momentum imparted to the nuclei is mediated by a virtual electronic transition. The efficiency of this process is scaled by the electronic susceptibility, where the signal amplitude follows 

\begin{equation}
    A_{ISRS} \propto \text{Im}[\chi^{(3)}] \approx \left(\frac{1}{\Gamma_{elec}}\right)^2
\end{equation}

where $\Gamma_{elec}$ is the homogeneous electronic linewidth ($1/T_2^*$). Hence, the amplitude of the ISRS modes is tied to the electronic dephasing time, $T_2$. As the temperature increases, phase coherence of the electronic polarization is rapidly eroded by thermally activated fluctuations in the organic-inorganic interface, thereby reducing the efficiency of the ISRS process before significant nuclear momentum can be established. Hence, the reduction of $M3$ and $M4$ is not merely a sign of vibrational damping, but a spectroscopic signature of the collapse of the electronic coherence under the influence of dynamic disorder. 

In contrast, the DECP modes exhibit an increase in relative intensity, implying that the displacive force actually becomes more effective at higher temperatures. In the DECP framework, the initial displacement $Q_0$ is defined by the shift in the PES minimum upon excitation ($\Delta$). Crucially, in 2D-MHPs with high static distortion (large BAV), the PES is highly anharmonic.

Within the density matrix formalism, the amplitude of the coherent phonon signal is determined by the initial nuclear displacement ($Q_0$) and the initial momentum ($P_0$). In our system, the enhancement of the excited state pathways at higher temperatures can be traced directly to the displacive force. The initial displacement ($Q_0$) in the DECP mechanism is defined by the shift in the PES minimum, quantified as:

\begin{equation}
    Q_0 = \Delta = \frac{\partial V_{ex}/\partial Q - \partial V_{gr}/\partial Q}{M\omega^2_{phonon}}
\end{equation}

where $\partial V_{ex}/\partial Q$ and $\partial V_{gr}/\partial Q$ represent the gradients of the excited and ground-state PESs at the Frank-Condon point, M is the effective mass of the oscillator, and $\omega_{ph}$ is the phonon frequency. While the ground-state gradient is relatively fixed by the equilibrium crystal structure, the excited-state gradient is highly sensitive to the local configurational landscape. We hypothesize that as temperature increases, the inorganic framework samples a broader range of octahedral distortions as indicated by the increased bond angle variance (Fig.~\ref{fig:Raman_temp}(h)). 

In the regime of high distortion, the exciton accesses regions of the PES where the excited-state gradient is significantly steeper relative to the ground state (Fig.~\ref{fig:Raman_temp}(g)\&(h)), resulting in a temperature-dependent renormalization of the effective displacement $\Delta$. Consequently, even as thermal dephasing suppresses the impulsive momentum-driven pathways ($P_0$), the displacive excitation ($Q_0$) is enhanced by the increased flexibility of the lattice. This transformation of the potential energy landscape ensures that population-driven structural reorganization becomes the dominant mechanism for phonon coherence at higher temperatures.  

\section{Conclusions}

In summary, we have unraveled the complex excitonic and vibrational dynamics in the chiral 2D-MHP \ce{(R-MBA)2PbI4} through a combination of ultrafast phase-resolved spectroscopy and temperature-dependent analysis. Our results establish that the dual excitonic resonances (XL and XH), with an unusually large energy separation ($\sim 90$ meV), originate from distinct electronic manifolds rather than simple phonon replicas or fine-structure effects.

Through phase-resolved RISRS, we have analytically disentangled the impulsive (ISRS) and displacive (DECP) mechanisms of lattice coherence. We identify $M3$ and $M4$ as ground-state coherences driven by impulsive momentum ($P_0$), while the remaining low-frequency modes are population-driven structural reorganizations initiated by a finite coordinate shift ($Q_0$) in the excited state. The observed $\pi$-phase inversion between $M1$ and $M2$ reveals a stereospecific coordinate-space relaxation characterized by simultaneous expansion and contraction of the inorganic lattice.

Most critically, the temperature-dependent behavior of these vibrational signatures shows a potential energy landscape that is not static, but rather undergoes a profound thermal renormalization. We propose a framework where rising temperature acts as a mechanistic filter, selectively suppressing or amplifying specific excitation pathways of coherent phonons. The relative amplitude reduction of the ISRS pathway (momentum-driven) demonstrates the thermal erosion of coherence driven by the increase of anharmonicity with temperature. Conversely, the increased lattice flexibility at elevated temperatures allows the system to escape its rigid low-temperature minima and sample the more anharmonic "softer" regions of the PES. In this scenario, the exciton experiences a steeper gradient in the excited-state relative to the ground-state. The resultant increase in the displacive driver ($Q_0$) effectively renormalizes the Huang-Rhys parameter, ensuring that even as impulsive momentum ($P_0$) reduces, the population-driven structural reorganization gains dominance.

These results demonstrate that the lattice flexibility inherent to these hybrid frameworks, which is enhanced due to the structural constraints imposed by the chiral organic, is not merely a source of spectral noise, but a structural determinant that reshapes the topology of the excitonic landscape governing how the lattice responds to light. By resolving the phase of the coherent vibrational response, we show that the transition from impulsive to displacive coupling is a deterministic consequence of the lattice's thermal flexibility. At low temperatures, the rigid framework constrains the system to ground-state Raman scattering; however, as temperature increases, the lattice 'softens,' allowing the electronic population to initiate a large, directional lattice reorganization.

\section{Author Contributions}
KAK performed the spectroscopy measurements and analyzed the data under the supervision of ARSK. MPH fabricated the samples and performed the crystallography analysis. The manuscript was written by KAK and ARSK with inputs from MCB and MPH. The project was conceived and coordinated by ARSK.  

\section{Supporting Information}
\begin{itemize}
    \item Sample Preparation
    \item Material Characterization 
    \item Ultrafast Transient Absorption Spectroscopy - Methods and additional details
    \item Resonant Impulsive Stimulated Raman Scattering  - Analysis
\end{itemize}
\section{Data Availability Statement}
The data underlying this study are openly available in Zenado at [10.5281/zenodo.20544512].
\section{Acknowledgements}

ARSK acknowledges funding from the National Science Foundation CAREER grant (CHE-2338663), start-up funds from Wake Forest University, funding from the Center for Functional Materials at Wake Forest University. Any opinions, findings, and conclusions or recommendations expressed in this material are those of the authors(s) and do not necessarily reflect the views of the National Science Foundation. KAK acknowledges the support of the U.S. Department of Energy, Office of Science, Office of Workforce Development for Teachers and Scientists, Office of Science Graduate Student Research (SCGSR) program. The SCGSR program is administered by the Oak Ridge Institute for Science and Education for the DOE under contract number DE-SC0014664.


\clearpage

\providecommand{\noopsort}[1]{}\providecommand{\singleletter}[1]{#1}%
\providecommand{\latin}[1]{#1}
\makeatletter
\providecommand{\doi}
  {\begingroup\let\do\@makeother\dospecials
  \catcode`\{=1 \catcode`\}=2 \doi@aux}
\providecommand{\doi@aux}[1]{\endgroup\texttt{#1}}
\makeatother
\providecommand*\mcitethebibliography{\thebibliography}
\csname @ifundefined\endcsname{endmcitethebibliography}  {\let\endmcitethebibliography\endthebibliography}{}



\clearpage
\begin{center}
    \noindent\large{\textbf{Supplementary Information: Temperature-Induced Crossover of Coherent Phonon Mechanisms in Chiral 2D Perovskites}} \\
  \vspace{0.5cm} 

  \normalsize
\end{center}

\section{Sample Preparation}

\subsection{Growth of \ce{(R/S-MBA)2PbI4} single crystals}
\ce{PbI4} (413 mg, 0.895 mmol, 99.99\% TCI) was dissolved in 5.5 mL of HI (57 wt\% in \ce{H2O}, Beantown Chemical) and 0.5 mL of \ce{H3PO2} (50 wt\% in \ce{H2O}, Sigma Aldrich) by heating above 100\degree C. The solution was then removed from heat and allowed to cool to room temperature before the dropwise addition of R/S-$\alpha$-methylbenzylamine (228 $\mu$L, 1.80 mmol, 98\% enantiomeric excess, Sigma Aldrich). The heated above 100\degree C until completely dissolved, followed by cooling to room temperature at a rate of -1\degree C/h. Orange needle like crystals here collected via vacuum filtration followed by rinsing thoroughly with diethylether (Oakwood chemical). The product was dried at 55\degree C under vacuum overnight. 

\subsection{Thin-films}
Approximately 1.5 × 1.5 cm$^2$ glass substrates were cleaned in UV-ozone for 15 minutes. In a glovebox, solutions of \ce{(R/S-MBA)2PbI4} in DMF (50 mg/mL) were spin coated onto the substrates for 20 s at 3000 RPM, followed by annealing for 5 minutes at 100\degree C. Film thickness was estimated to be 20-40 nm from measurements on a dektak profileometer.

\section{Sample Characterization}
\subsection{Low-Temperature Linear Absorption}
The linear absorption for all three sample variations is presented below. The origin of the energy shift observed in the racemic mixture is outside the scope of this manuscript. However, it is clear that the two-peak structure at low temperature is present in all three variations. 

\setcounter{figure}{0}
\renewcommand{\figurename}{FIG.}
\renewcommand{\thefigure}{S\arabic{figure}}

\begin{figure}[H]
    \centering
    \includegraphics[width=10cm]{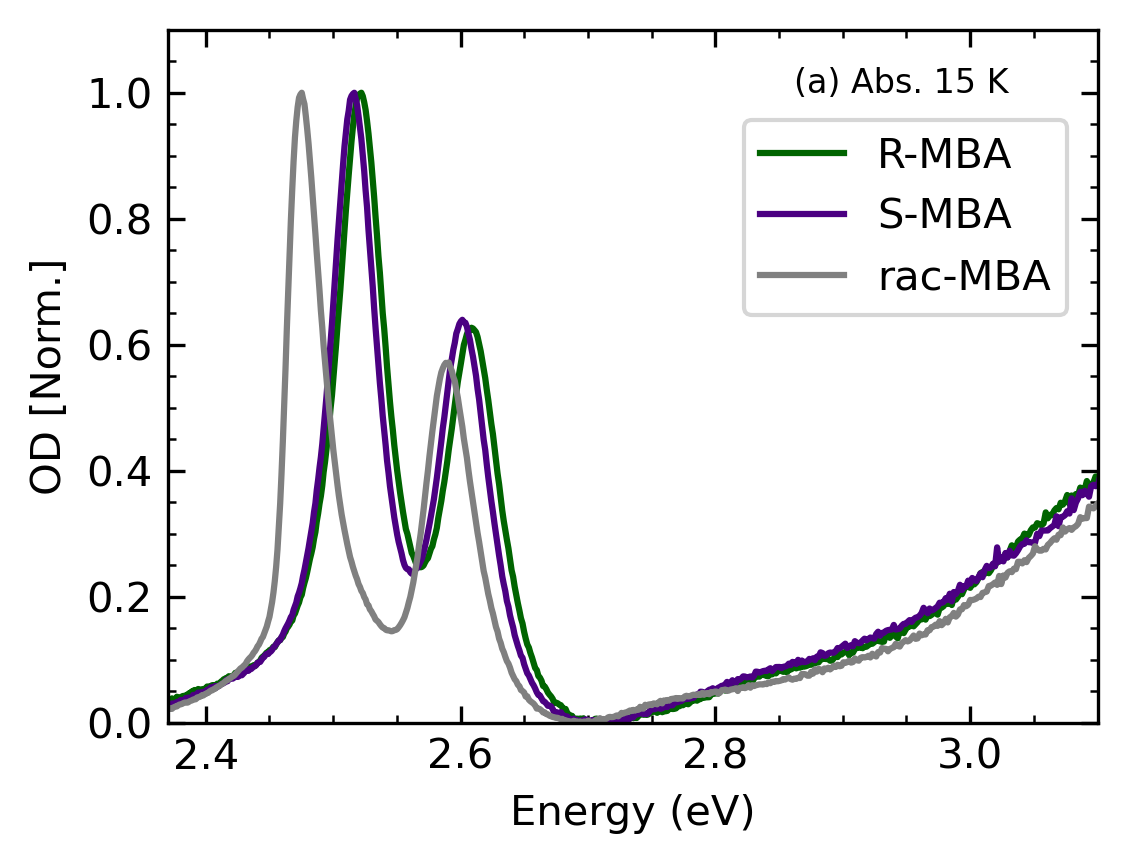}
    \caption{Linear absorption spectra measured at 15\,K for \ce(R/S/rac-MBA).} 
    \label{fig:rac_Abs}
\end{figure}

\subsection{Crystallography}
The bond angle variance (BAV), $\sigma^2$, was determined for \ce{(R/S/rac-MBA)PbI4}, using the crystallographic information files (CIF), reported in Ref.~\citenum{jana2020organic}, and VESTA, a 3D visualization program for structural models. 

\begin{figure}[H]
    \centering
    \includegraphics[width=10cm]{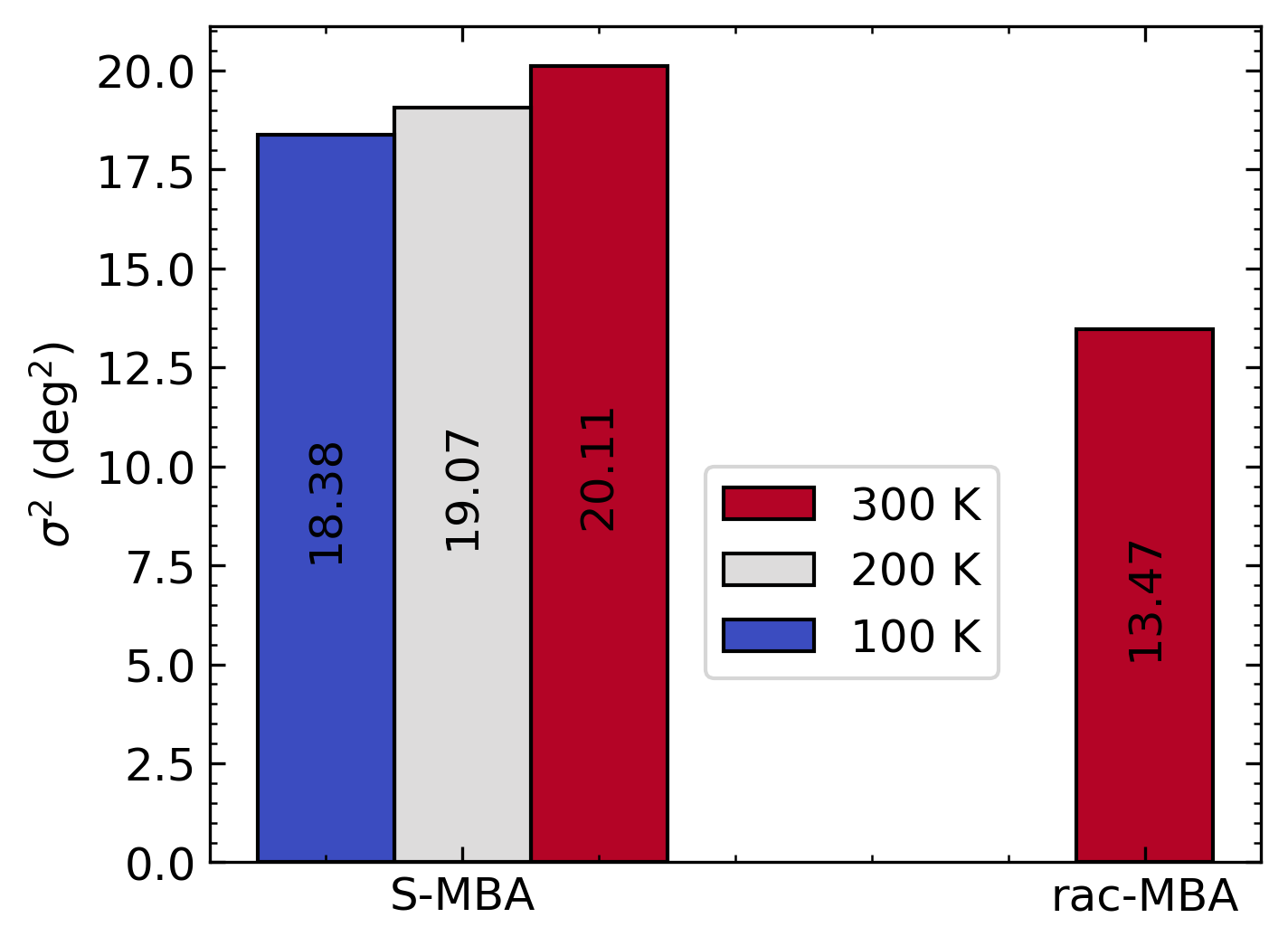}
    \caption{Bond angle variance ($\sigma^2$) determined for \ce{(S-MBA)2PbI4} and the corresponding racemic mixture, denoted as S-MBA and rac-MBA, respectively.} 
    \label{fig:crystal_params}
\end{figure}

\subsection{X-ray Diffraction (XRD)}
Thin-film X-ray diffraction patterns were collected with a Rigaku SmartLab diffractometer equipped with a \ce{Cu} K$\alpha$ source ($\lambda$ = 1.54059 \AA) in the $\theta$-2$\theta$ Bragg Brentano geometry with a 0.05\degree\ step size and 2 second/step scan rate. 

\begin{figure}[H]
    \centering
    \includegraphics[width=10cm]{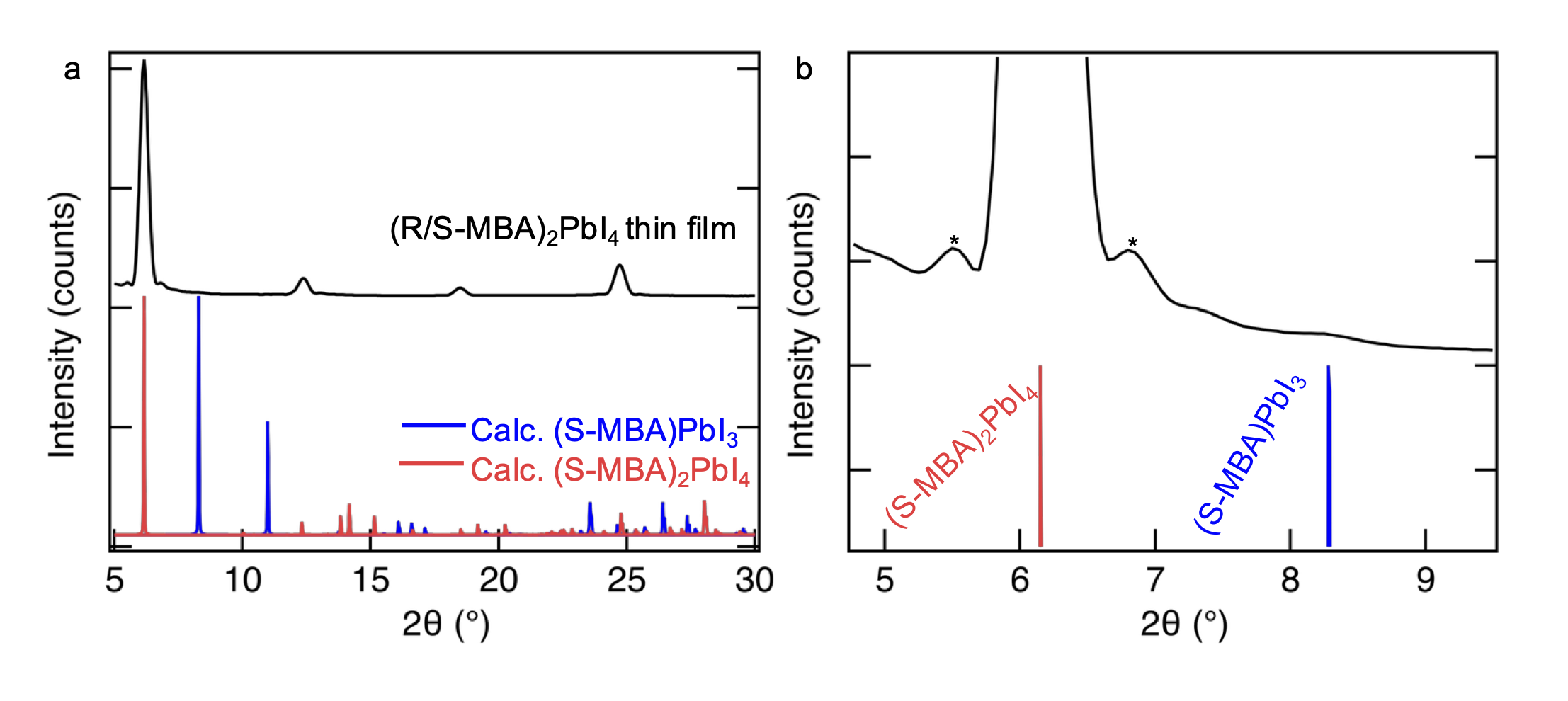}
    \caption{XRD of the \ce{(S-MBA)2PbI4} thin films. (a) XRD pattern showing oriented thin films (inorganic planes parallel to the substrate) compared to calculated 2D \ce{(S-MBA)2PbI4} and 1D \ce{(S-MBA)PbI3}. (b) Zoomed in view showing minimal contributions in the film from a 1D phase (\ce{(S-MBA)PbI3}), not likely to feature prominent absorbance in the film during spectroscopy experiments. The * shows satellite peaks likely corresponding to thickness fringes from the film.} 
    \label{fig:XRD}
\end{figure}

\newpage

\section{Ultrafast Transient Absorption Spectroscopy}
Transient absorption spectroscopy measurements were performed using a pulsed femtosecond laser (Pharos Model PH1-20-0200-02-12, Light Conversion) emitting 1030\,nm pulses at 1\,kHz with a pulse duration of $\sim$200\,fs. The pump beam (2.48\,eV-3.10\,eV) was generated by feeding half of the laser output to a commercial optical parametric amplifier (Orpheus, Light Conversion) while 5\,mW was focused onto a sapphire crystal to obtain a single filament white-light continuum covering the spectral range 450\,nm - 700\,nm for the probe beam. The probe beam transmitted through the sample was detected by a high-speed camera (FLC 3030, EB Stressing) in combination with a high-resolution spectrometer (SpectraPro SP-2150, Princeton Instruments). All measurements were carried out in a vibration-free closed-cycle cryostation (Montana Instruments). 

\subsection{Spectral Lineshape Analysis}\label{sec:TA_lineshape}

\begin{figure}
    \centering
    \includegraphics[width=\columnwidth]{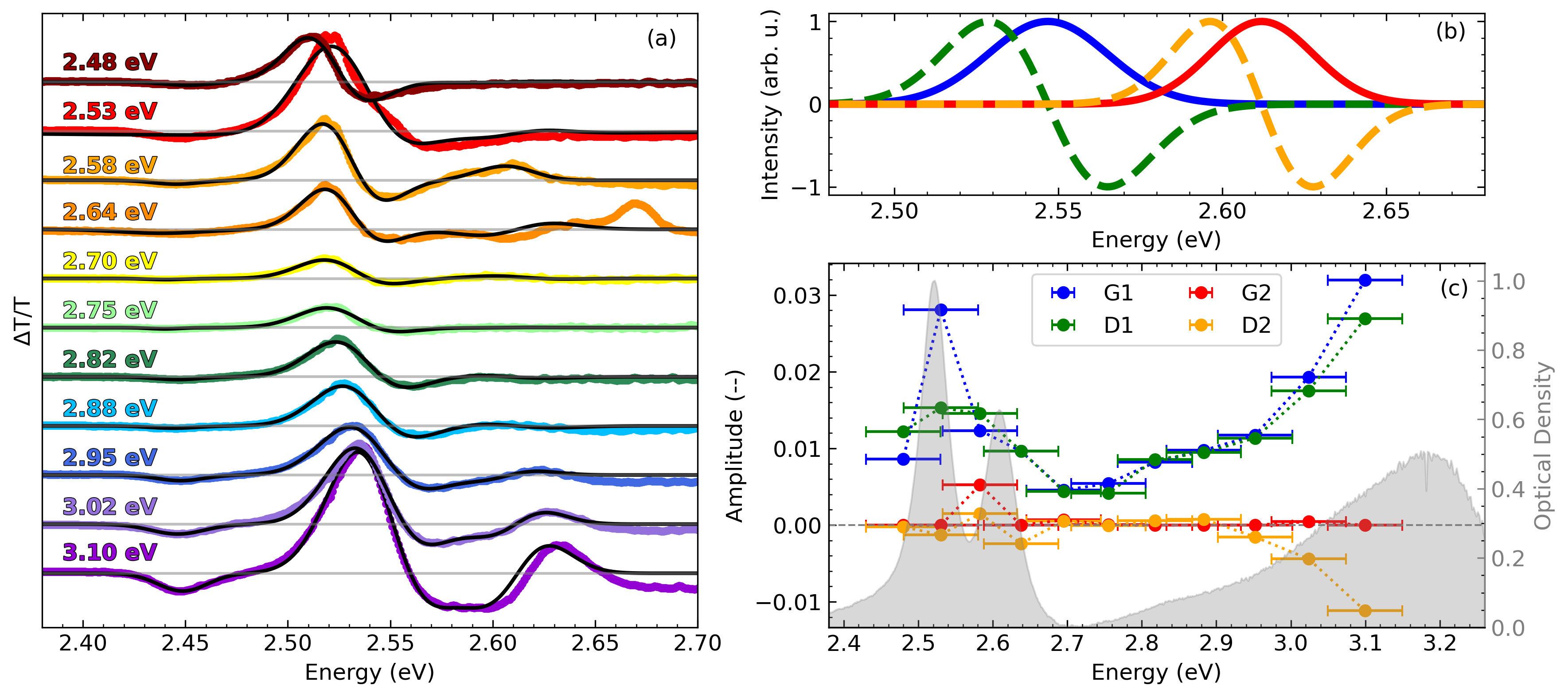}
    \caption{(a) TA cuts (colorful lines) taken at 3\,ps for various pump energies and their corresponding fits (black lines). For the TA fits, we assumed the spectra were composed of two Gaussians and their corresponding first derivatives. (b) Visual representation of the components assumed for the TA fitting, where G1 (blue) and G2 (red) are Gaussian peaks and D1 (green) and D2 (yellow) are their corresponding first derivatives. (c) The relative amplitudes of the fitting components, mapped onto the linear absorption spectra of \ce{(R-MBA)2PbI4} (gray shaded) measured at 15\,K.}
    \label{fig:PP_cuts_fits}
\end{figure}

To further analyze the TA spectra, we take spectral cuts at a fixed delay of 3\,ps across a range of pump energies. Each resulting lineshape is fit using a combination of two Gaussian functions -- G1 and G2, corresponding to the ground-state bleach features -- and their respective first derivatives, D1 and D2, which capture the derivative-like photoinduced absorption feature (Figs.~\ref{fig:PP_cuts_fits}(a)–(c)). When the amplitudes of these components are mapped onto the linear absorption spectrum (Fig.~\ref{fig:PP_cuts_fits}(c)), it becomes evident that, in general, contributions from the XL resonance primarily govern the TA signal.

At all pump energies, the spectral response at the XL energy can be reproduced by the combination of G1 and D1, whose amplitudes closely track the optical density profile, indicating that XL is efficiently photo-generated across a range of excitation energies. At the energy of XH resonance, however, the lineshape is dominated by the derivative lineshape (D2), with the ground-state bleach (G2) contribution only present when the pump energy is tuned resonantly to the XH transition. This suggests that the XH state is populated only under resonant excitation, although its energy is modulated in the presence of the photo-excited population generated with other pump energies that are resonant with XL and the continuum.

We also note that opposite signs for the D1 and D2 contributions, which indicate a photo-induced blue shift in the XL energy, and a red shift in the XH energy. This asymmetry in the photo-induced energy shifts — a blue shift for XL and a red shift for XH — points to fundamentally different physical mechanisms governing the behavior of each excitonic transition. Given that XL and XH are spectrally and dynamically distinct, we consider a scenario in which they originate from transitions into different regions of the carrier reciprocal space, such as distinct conduction-band minima or valleys (Fig.~\ref{supp-fig:energy_schematic}). In this picture, XL and XH excitons are associated with electronic states of differing orbital character, effective mass, and coupling to the lattice, which in turn modulate their response to photoexcitation. 

\begin{figure}
    \centering
    \includegraphics[width=\columnwidth]{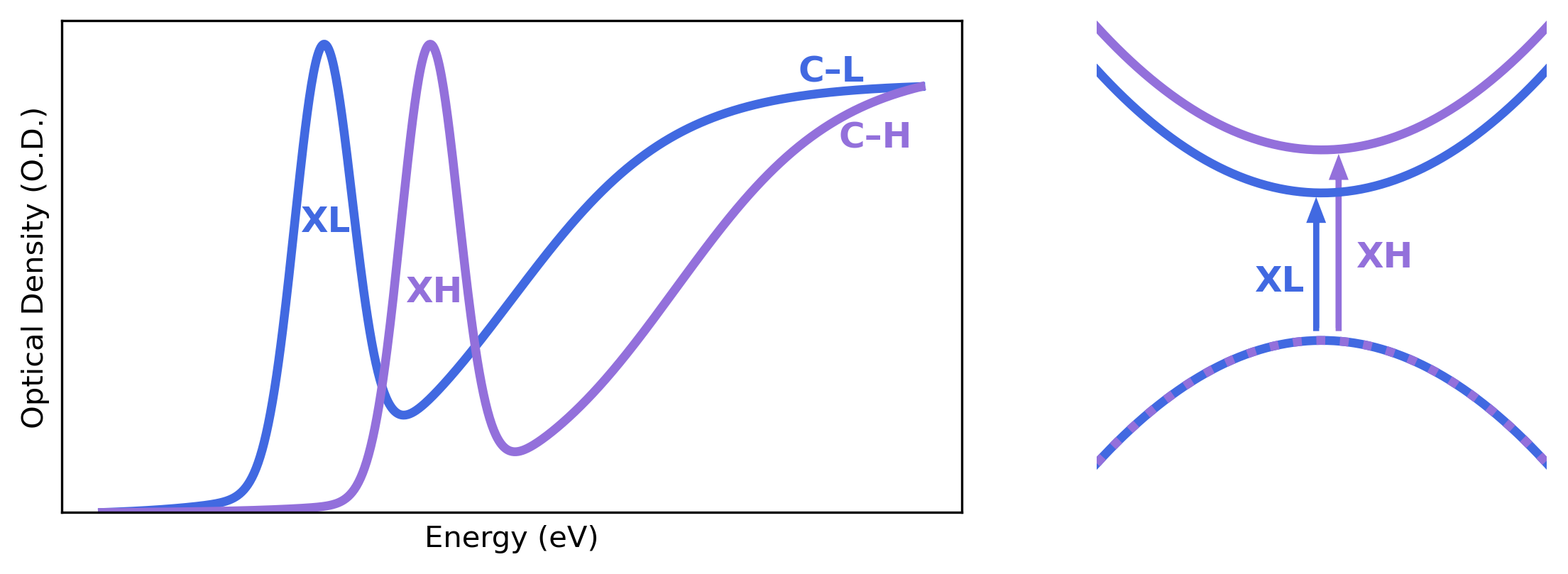}
    \caption{Schematic illustration of a possible explanation for the excitonic absorption features observed in \ce{(R/S-MBA)2PbI4}.}
    \label{fig:energy_schematic}
\end{figure}

The blue shift of XL is consistent with phase-space filling and Coulomb screening effects in the populated state. Upon photoexcitation, carriers fill states near the XL conduction-band minimum, blocking further transitions and reducing exciton binding energy. These effects combine to push the XL resonance to higher energy~\cite{schmitt1985theory, haug1985basic, haug1994quantum}. In contrast, the red shift of XH — despite the absence of direct population — suggests that its energy is renormalized indirectly, possibly through bandgap renormalization or a transient Stark effect induced by carrier accumulation in the XL-associated state~\cite{liu2022charge, trinh2015many, saran2017giant}. This nonlocal modulation implies that the XH transition is sensitive to the electrostatic environment generated by photoexcited carriers elsewhere in the Brillouin zone.

In the simple scenario shown in Fig.~\ref{supp-fig:energy_schematic}, the photo-excitation in our experiment generates holes in the valence band and electrons in the lower lying conduction band. Importantly, there are no carriers generated in the conduction band associated with the XH transition. The presence of the holes in the valence band results in the renormalization of the bandgap, thus red-shifting the XH. Note that in this hypothetical scenario, XH will have a larger exciton binding energy even though it is blue-shifted with respect to XL. The continuum associated with XH lies at much higher energy, beyond the pump energies used in our experiment. If XL and XH transitions correspond to different valleys in a multi-valley band structure, the observed shifts reflect valley-specific many-body interactions and screening dynamics.

While derivative-like lineshapes in the transient spectra could, in principle, be attributed to lattice heating effects arising from high excitation densities, we rule out this mechanism as the primary origin of the observed spectral shifts. This conclusion is based on the fact that these derivative features appear within 200\,fs of photoexcitation, limited by the experimental resolution, and significantly shorter than the thermalization timescale typically associated with lattice heating~\cite{cushing2018hot, tkachenko2016transient, ziaja2015time}. Nonetheless, this does not imply that lattice heating is absent in the experiment. 

\begin{figure}
    \centering
    \includegraphics[width=\columnwidth]{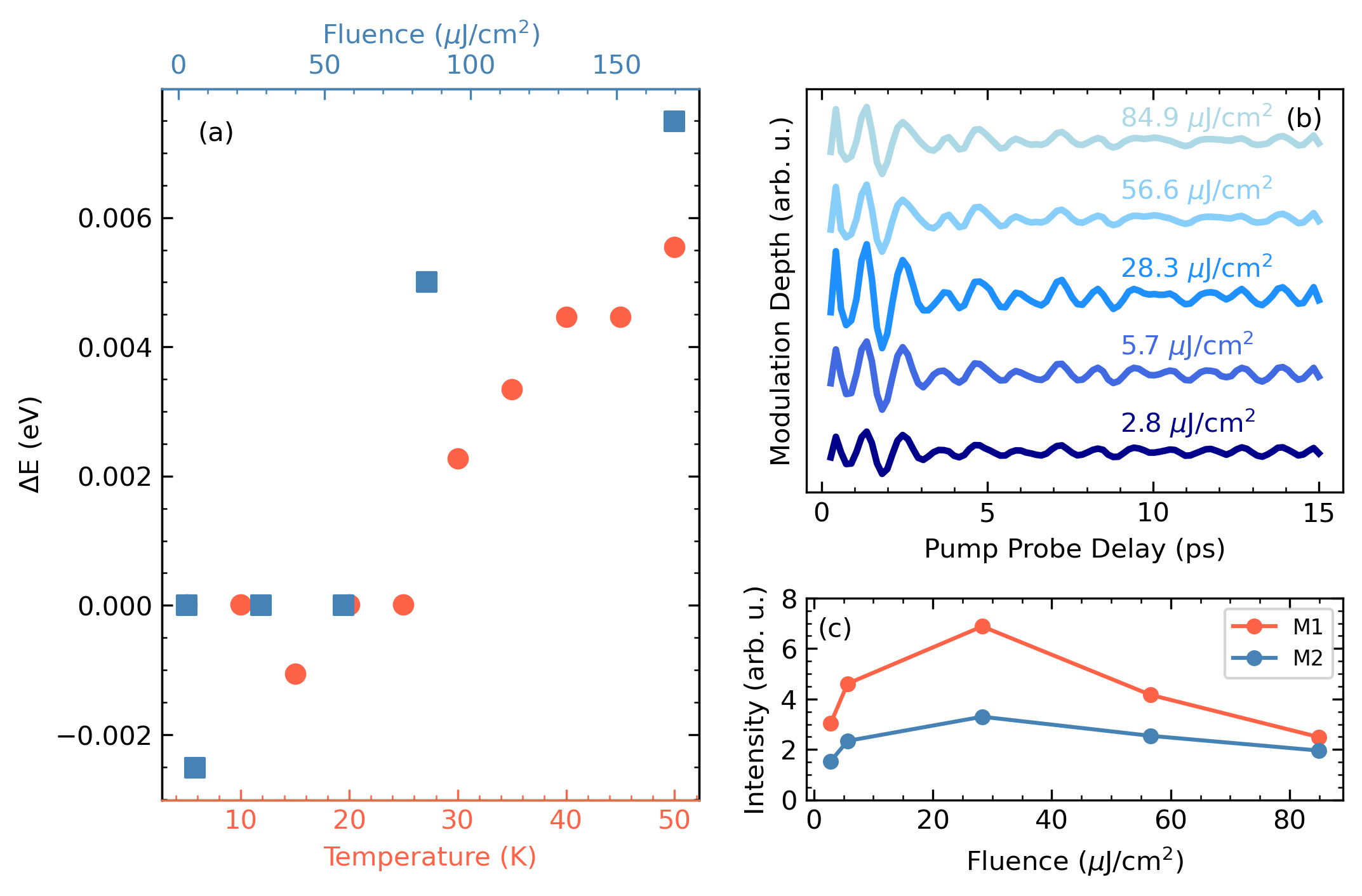}
    \caption{(a) Energy shift as a function of fluence and temperature. $\Delta$E as a function of fluence was calculated by tracking the position of the ground state bleach in the TA data. The temperature-dependent energy shift was determined through the peak shifts in the temperature-dependent linear absorption spectra. (b) Fluence-dependent modulation depth due to phonons as a function of pump-probe delay, measured using the resonant impulsive stimulated Raman scattering (RISRS) technique. (c) Raman intensities as a function of pump power for phonon modes $M1$ (3.8\,eV) and $M2$ (5.2\,eV).}
    \label{fig:Lattice_Heating}
\end{figure}

Indeed, we do identify the onset of steady-state heating at elevated temperatures. In order to determine the impact of increasing pump fluence on the lattice temperature, we compared the energy shift as a function of pump fluence and temperature. The former was determined by tracking the peak position of the ground state bleach (GSB) from transient absorption, while the latter was determined through the peak shift of the XL resonance in the temperature-dependent linear absorption spectra. 

As shown in Fig.~\ref{supp-fig:Lattice_Heating}(a), the energy of the G1 feature (blue squares) remains stable at fluences below 50\,$\mu$J/cm$^2$, with minor deviations due to experimental noise. Beyond this threshold, a pronounced and monotonic blue shift is observed. By comparing this fluence-dependent shift to the temperature-dependent linear absorption data (orange circles in Fig.~\ref{supp-fig:Lattice_Heating}(a)), we estimate that high-fluence excitation induces a local steady-state temperature rise exceeding 50 K, despite a nominal cryostat setpoint of 5 K. 

Further evidence of heating is found in the coherent phonon dynamics, shown in Fig.~\ref{supp-fig:Lattice_Heating}(b). These oscillatory features, arising from impulsive excitation of Raman-active phonons (discussed extensively in the main text), exhibit a fluence-dependent behavior: their amplitude increases with fluence up to 30\,$\mu$J/cm$^2$, beyond which it begins to diminish. This turnover coincides with the fluence threshold where the G1 blue shift becomes prominent, reinforcing the interpretation that steady-state lattice heating modifies both the spectral and dynamical lattice response of the system. To mitigate these thermal effects and preserve the fidelity of the ultrafast measurements, we restrict our analysis to relatively low-fluence regimes where lattice heating remains minimal.

\newpage{}

\section{Resonant Impulsive Raman Scattering}\label{sec:RISRS_Data}
\subsection{RISRS Data Processing}
The electronic dynamics are subtracted from the differential transmission spectra using a high-order polynomial fit. The resulting modulated response is then fast Fourier transformed at each probe-energy to obtain a beating map. The Raman spectra shown in the text are obtained by integrating over the relevant probe energies. For a more detailed explanation of this process, see Ref.~\citenum{thouin2019phonon} and Ref.~\citenum{koch2025structure}. 
\begin{figure}[H]
    \centering
    \includegraphics[width=6cm]{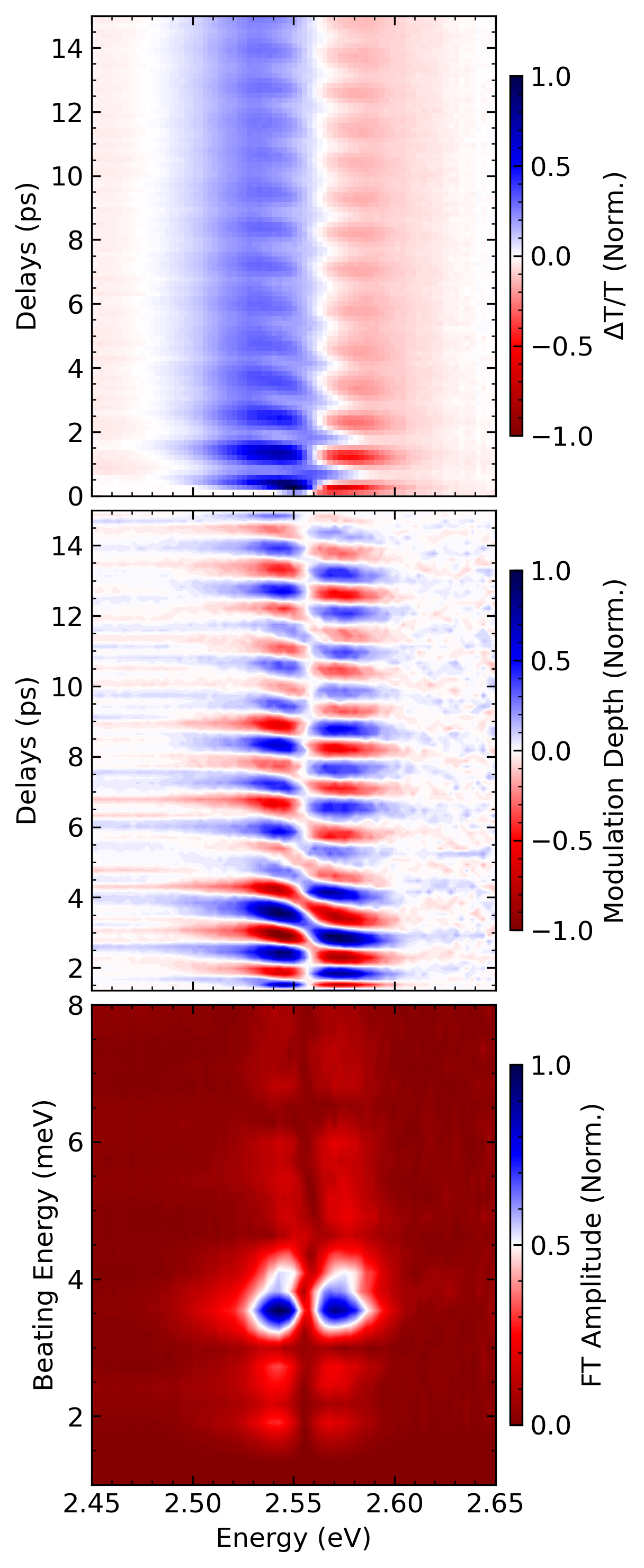}
    \caption{Demonstration of RISRS data processing.}
    \label{fig:RISRS_data_processing}
\end{figure}

\subsection{Phase Profile}
Further analysis of the real and imaginary components resulting from the Fourier Transform allows the extraction of the phase profile of the coherent phonon modes. The relative phase between modes is useful for determining the excitation pathways of the phonon modes (ground vs. excited state mode). The phase profile extracted from the Fourier Transform is shown below. The black dashed lines correspond to the beating energy line cuts shown in Figure 3 of the main text.

\begin{figure}[H]
    \centering
    \includegraphics[width=\columnwidth]{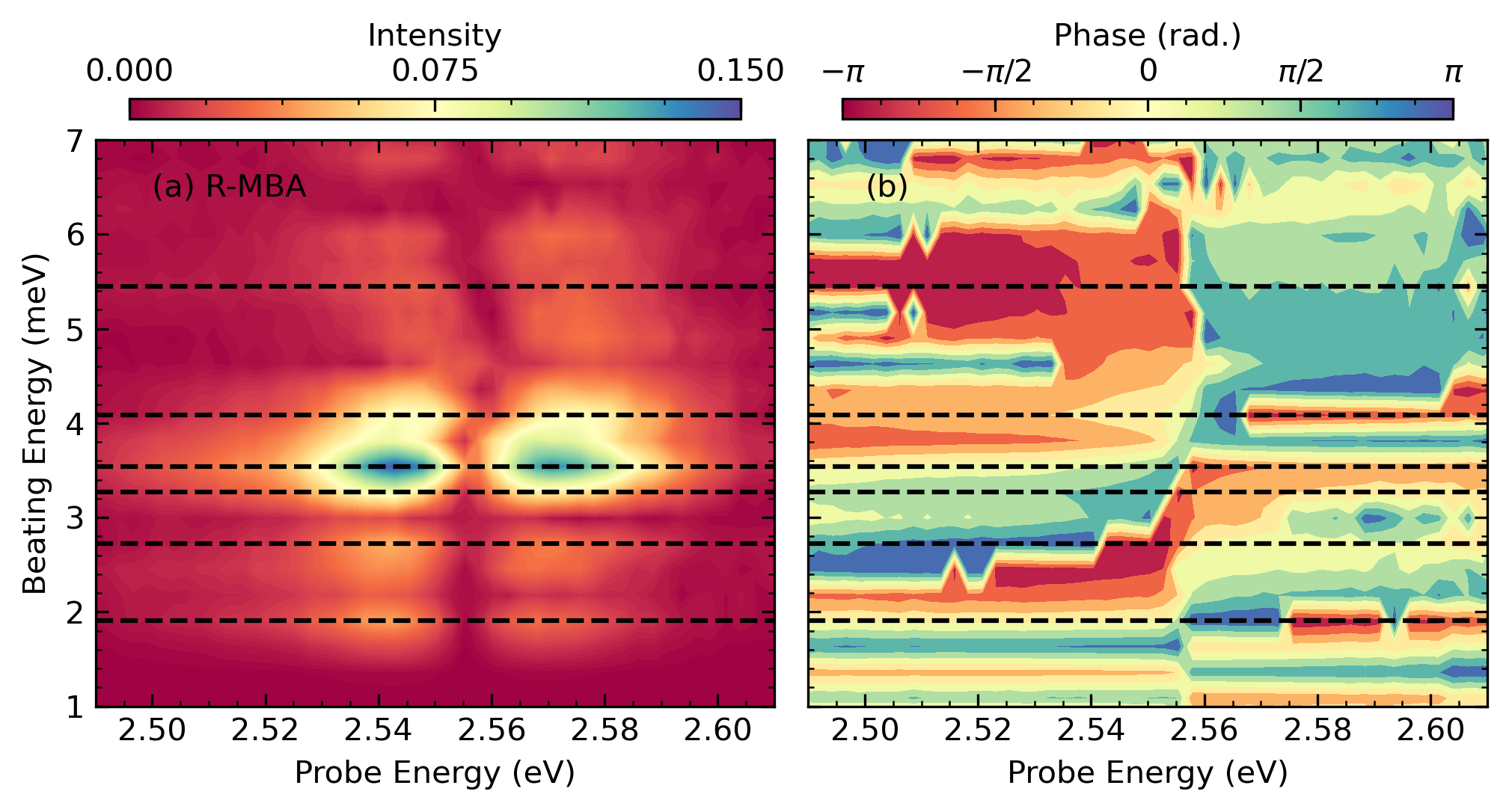}
    \caption{(a) Beating map measured at 5\,K, with a 2.53\,eV pump energy, and (b) its corresponding phase map, extracted from the real and imaginary components of the Fourier Transform. The black dashed lines correspond to the specific phonon energy line cuts shown in Figure 4 of the main text. }
    \label{fig:SI_Raman_Phase}
\end{figure}

\newpage
In the main portion of the article, the phase is reported for R-MBA, but as shown below, the phase profile is similar for the corresponding chiral enantiomer, S-MBA.

\begin{figure}[h]
    \centering
    \includegraphics[width=7cm]{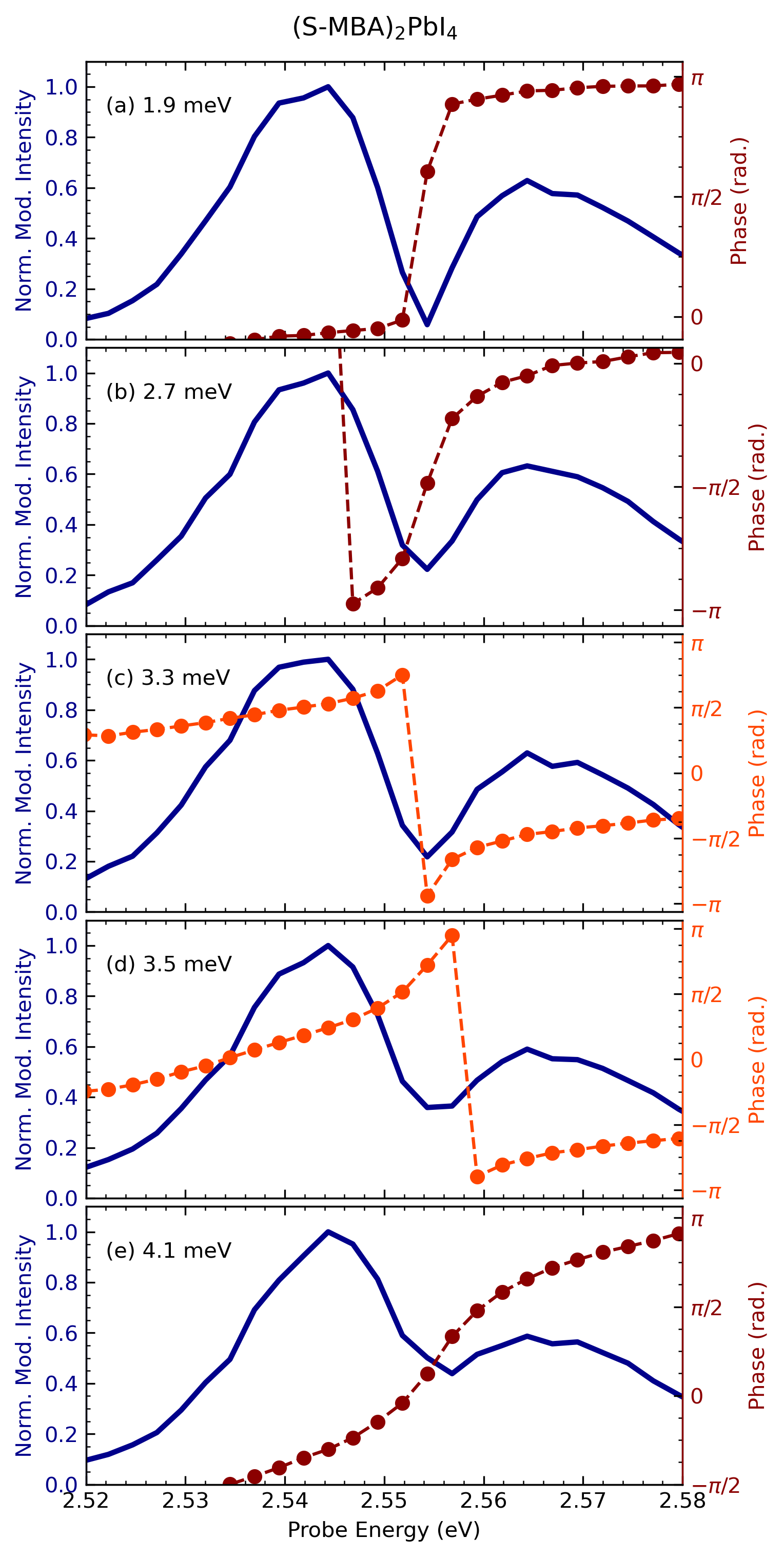}
    \caption{(a)-(e) The modulation spectra (dark blue line) and corresponding phase profiles (orange (dark red) circles for ISRS (DECP) modes) of the various phonon modes observed in the beating map measured at 5\,K with a pump energy of 2.53\,eV of \ce{(S-MBA)2PbI4}.} 
    \label{fig:Raman_phase}
\end{figure}

\subsection{Vibrational Coherence Simulations}\label{sec:Kumar_simulations}

\subsubsection*{1. Theoretical Framework \& Complex Susceptibility}
To model the spectrally resolved pump-probe amplitude and phase profiles across the electronic resonance window, we employ the vibronic model pioneered by Champion and co-workers. This approach evaluates the differential transmission or absorption by treating the coherent phonon as a macroscopic nuclear wavepacket that modulates the material's complex linear susceptibility, $\chi^{(1)}(E_{pr})$.

The unperturbed electronic transition is modeled as a complex Lorentzian resonance centered at energy $E_0$ with a homogeneous electronic dephasing linewidth $\Gamma$:
\begin{equation}
\chi^{(1)}(E_{pr}) = \frac{\mu^2}{(E_0 - E_{pr}) - i\Gamma}
\end{equation}
where $\mu$ is the electronic transition dipole matrix element and $E_{pr}$ is the probe photon energy. 

Thermal effects and vibrational dephasing are incorporated via a time-domain dephasing function $g(s)$, which tracks the evolution of the system over the coherence time variable $s$:
\begin{equation}
g(s) = \frac{\Delta_1^2}{2} \left[ \coth\left(\frac{\hbar \omega_0}{2k_B T}\right) (1 - \cos(\omega_0 s)) + i \sin(\omega_0 s) \right]
\end{equation}
where $\Delta_1$ represents the dimensionless electron-phonon coupling constant (displacement of the potential energy surface along the normal mode coordinate), $\omega_0$ is the ground-state phonon frequency, and $T$ is the temperature. The ground- and excited-state nuclear coherence half-responses, $K_g(s)$ and $K_e(s)$, are expressed as:
\begin{equation}
K_g(s) = \exp\left( -\frac{i E_0 s}{\hbar} - \frac{\Gamma |s|}{\hbar} \right) \exp(-g(s))
\end{equation}
\begin{equation}
K_e(s) = \exp\left( +\frac{i E_0 s}{\hbar} - \frac{\Gamma |s|}{\hbar} \right) \exp(-g(s))
\end{equation}
Taking the half-Fourier transforms of these coherence functions yields the fundamental ground-state ($\Phi$) and excited-state ($\Theta$) spectral response functions:
\begin{equation}
\Phi(E_{pr}) = i \int_0^\infty \exp\left(\frac{i E_{pr} s}{\hbar}\right) K_g(s) \, ds
\end{equation}
\begin{equation}
\Theta(E_{pr}) = i \int_0^\infty \exp\left(\frac{i E_{pr} s}{\hbar}\right) K_e^*(s) \, ds
\end{equation}

\subsubsection*{2. Wavepacket Generation \& Temporal Phase Orthogonality (ISRS vs. DECP)}
The interaction of an ultrashort pump pulse with the system projects the ground-state nuclear wavepacket into initial phase-space coordinates designated by a coordinate displacement $Q_{g0}$ and an impulsive momentum ``kick'' $P_{g0}$. These initial moments are evaluated numerically by taking the convolution of the probe pulse envelope function, $G(E) = \exp\left[-\frac{\tau_p^2}{8\ln 2} \left(\frac{E - E_c}{\hbar}\right)^2\right]$ (centered at probe energy $E_c$ with duration $\tau_p$), with the imaginary and real parts of the ground spectral response:
\begin{equation}
Q_{g0} = \mathcal{A}_0 \int G(E - E_c) G(E - E_c - \hbar\omega_0) \left[ \text{Im}\,\Phi(E) - \text{Im}\,\Phi(E - \hbar\omega_0) \right] dE
\end{equation}
\begin{equation}
P_{g0} = \mathcal{A}_0 \int G(E - E_c) G(E - E_c - \hbar\omega_0) \left[ \text{Re}\,\Phi(E) - \text{Re}\,\Phi(E - \hbar\omega_0) \right] dE
\end{equation}
where $\mathcal{A}_0$ collects pump-intensity scaling terms and the thermal occupation factor $\left[\coth\left(\frac{\hbar\omega_0}{2k_B T}\right) - 1\right]$. 

The relative phase shift between ISRS and DECP mechanisms originates purely from these temporal initial boundary conditions at $t=0$:
\begin{itemize}
    \item \textbf{Impulsive Stimulated Raman Scattering (ISRS):} Driven off-resonantly via a virtual electronic transition, the pump pulse acts as an instantaneous force impulse. Because the pulse duration is significantly shorter than the phonon period ($\tau_p \ll T_{\text{phonon}}$), the field acts as a delta-function impulse. At $t=0$, the physical displacement is zero ($Q_{g0} = 0$) while the momentum transfer is maximized ($P_{g0} \gg 0$). The initial temporal phase is given by:
    \begin{equation}
    p_g = -\arctan2(P_{g0}, Q_{g0}) \approx -\frac{\pi}{2}
    \end{equation}
    This boundary condition mandates a pure \textit{sine-like} initial trajectory, $Q_{\text{ISRS}}(t) \propto \sin(\omega_0 t)e^{-\gamma t}$.
    
    \item \textbf{Displacive Excitation of Coherent Phonons (DECP):} Driven under resonant conditions, real electronic absorption transfers population to the excited-state manifold, abruptly shifting the structural minimum of the potential energy surface by a distance $\Delta_1$. At $t=0$, the nuclear framework remains momentarily stationary but is instantly displaced relative to the new coordinate basin ($Q_{e0} \propto \Delta_1, P_{e0} = 0$). The initial temporal phase is thus pinned at:
    \begin{equation}
    p_e = 0
    \end{equation}
    This mandates a pure \textit{cosine-like} structural relaxation, $Q_{\text{DECP}}(t) \propto \Delta_1[1 - \cos(\omega_0 t)]e^{-\gamma t}$.
\end{itemize}
The fundamental quarter-period offset between the sine-driven impulse channel and the cosine-driven displacive channel establishes the fixed $\pi/2$ temporal phase orthogonality. When evaluated across the energy grid, this baseline phase offset shifts the entire phase profile vertically by $\pi/2$ relative to one another.

\subsubsection*{3. Spectrally Resolved Bessel Expansion \& Resonance Phase Jumps}
The total differential signal detected at a given probe energy is modeled using a sum-over-states expansion weighted by ordinary Bessel functions of the first kind ($J_n$). This non-perturbative framework accounts for multi-quantum transitions driven by the large-amplitude coherent state. The complex spectral components for the ground state ($C_g, S_g$) and excited state ($C_e, S_e$) are calculated via:
\begin{multline}
C_{g/e}(E_{pr}) = \frac{e E_{pr}}{\hbar} \sum_{n=-20}^{20} J_n(A_{g/e}\Delta_1) J_{n-1}(A_{g/e}\Delta_1) \times \\ 
\left[ G_- \cdot \text{Im}\,\Psi(E_{pr} + n\hbar\omega_0) + G_+ \cdot \text{Im}\,\Psi(E_{pr} + (n-1)\hbar\omega_0) \right]
\end{multline}
\begin{multline}
S_{g/e}(E_{pr}) = \frac{e E_{pr}}{\hbar} \sum_{n=-20}^{20} J_n(A_{g/e}\Delta_1) J_{n-1}(A_{g/e}\Delta_1) \times \\ 
\left[ G_- \cdot \text{Re}\,\Psi(E_{pr} + n\hbar\omega_0) - G_+ \cdot \text{Re}\,\Psi(E_{pr} + (n-1)\hbar\omega_0) \right]
\end{multline}
where $A_g = \sqrt{Q_{g0}^2 + P_{g0}^2}$, $A_e = |Q_{e0}|$, $G_{\pm} = G(E_{pr} - E_c)G(E_{pr} - E_c \pm \hbar\omega_0)$, and $\Psi$ represents $\Phi$ or $\Theta$ respectively. The final complex signals combining both temporal and spectral phase contributions are assembled as:
\begin{equation}
\mathcal{S}_{ISRS}(E_{pr}) = [C_g(E_{pr}) + iS_g(E_{pr})] \cdot e^{i p_g}
\end{equation}
\begin{equation}
\mathcal{S}_{DECP}(E_{pr}) = [C_e(E_{pr}) + iS_e(E_{pr})] \cdot e^{i p_e}
\end{equation}
The spectrally resolved intensity profile is given by $I(E_{pr}) = |\mathcal{S}(E_{pr})|$, and the total phase profile across the probe axis is extracted via $\phi(E_{pr}) = \text{unwrap}(\arg[\mathcal{S}(E_{pr})])$.

The characteristic sharp $\pi$ phase jump observed at the center of the absorption line is a purely spectral readout effect. Because the underlying susceptibility $\chi^{(1)}$ contains a resonant denominator, its derivative with respect to the nuclear displacement ($\partial\chi^{(1)}/\partial Q$) undergoes a zero-crossing node at the exact center of the electronic transition ($E_{pr} = E_0$). 
When sweeping from below resonance ($E_{pr} < E_0$) to above resonance ($E_{pr} > E_0$), the sign of this derivative inverts. This forces the calculated complex spectral response vectors to change signs simultaneously, $(C, S) \rightarrow (-C, -S)$. In complex polar coordinates, inverting the direction of a vector is equivalent to an instantaneous coordinate rotation of exactly $\pi$ radians ($180^\circ$). This creates the sharp phase discontinuities and localized intensity zeros observed at the central resonance node.

\subsubsection*{4. Structural Phase Shifts: Structural Inversion ($\Delta_1 > 0$ vs. $\Delta_1 < 0$)}
When comparing modes that correspond to opposing structural pathways along the lattice coordinates—such as a lattice expansion mode ($\Delta_1 > 0$) versus a lattice contraction mode ($\Delta_1 < 0$)—the phase profiles reveal a flat, parallel separation of exactly $\pi$ radians across the entire probe window.

This uniform phase shift is explicitly derived from the mathematical parity properties of ordinary Bessel functions under sign inversion, where $J_n(-x) = (-1)^n J_n(x)$. When evaluating the adjacent-order multi-quantum product terms within the summation loops, the sign inversion modifies the arguments as follows:
\begin{align}
J_n(-\Delta_1 A) \cdot J_{n-1}(-\Delta_1 A) &= (-1)^n J_n(\Delta_1 A) \cdot (-1)^{n-1} J_{n-1}(\Delta_1 A) \nonumber \\
&= (-1)^{2n-1} \left[ J_n(\Delta_1 A) J_{n-1}(\Delta_1 A) \right] \nonumber \\
&= -1 \cdot \left[ J_n(\Delta_1 A) J_{n-1}(\Delta_1 A) \right]
\end{align}
Because this negative sign is generated uniformly for every term in the sum-over-states loop, the global sign of the final complex spectral arrays is inverted:
\begin{equation}
\mathcal{S}_{\Delta_1 < 0}(E_{pr}) = -\mathcal{S}_{\Delta_1 > 0}(E_{pr})
\end{equation}
In the complex plane, a global multiplication by $-1$ corresponds to an absolute phase rotation of $e^{i\pi}$ applied evenly across the entire grid. Unlike the localized electronic resonance jump, this structural phase offset is uniform and independent of the probe photon energy, providing direct, unambiguous confirmation of opposing geometric displacement vectors.

\subsection{Temperature Dependent Modulation}
As discussed in the main text, there is a temperature-dependent evolution of the modulation lineshape. The amplitude of the modulation on the higher-energy side of the node decreases with increasing temperature, suggesting a temperature-dependent evolution of exciton-lattice coupling and the Huang-Rhys factor. The evolution of this lineshape is depicted in the figure below. 

\begin{figure}[H]
    \centering
    \includegraphics[width=9cm]{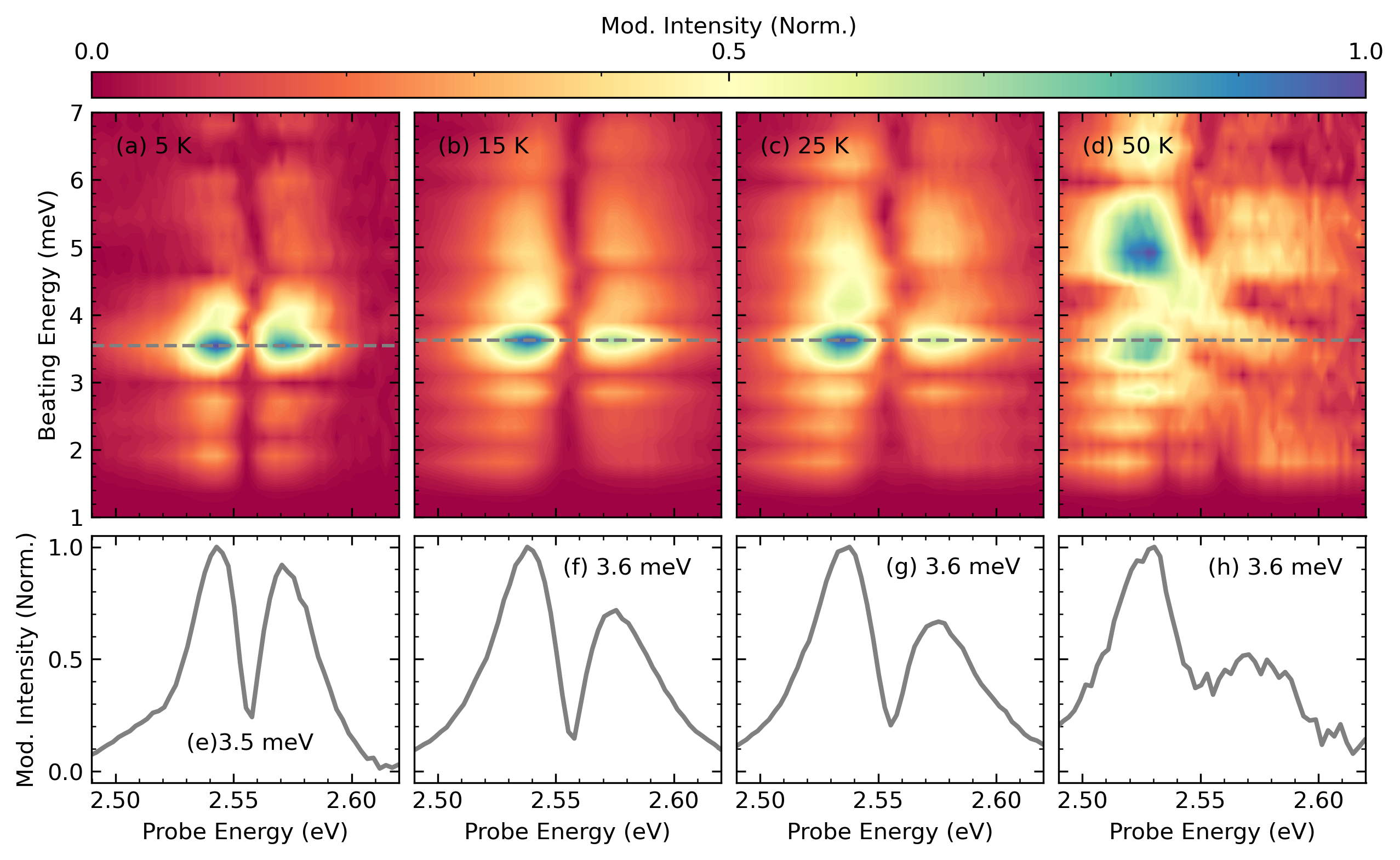}
    \caption{Temperature-dependent beating maps (a)-(d) and corresponding modulation spectra (e)-(h) of the main phonon mode. The modulation spectra are obtained via line cuts at the beating energy marked by the gray dashed line. The pump energy for this data set was 2.53\,eV, resonant with the XL transition.}
    \label{fig:SI_Mod_Temp_Dep}
\end{figure}

\newpage

\providecommand{\noopsort}[1]{}\providecommand{\singleletter}[1]{#1}%
\providecommand{\latin}[1]{#1}
\makeatletter
\providecommand{\doi}
  {\begingroup\let\do\@makeother\dospecials
  \catcode`\{=1 \catcode`\}=2 \doi@aux}
\providecommand{\doi@aux}[1]{\endgroup\texttt{#1}}
\makeatother
\providecommand*\mcitethebibliography{\thebibliography}
\csname @ifundefined\endcsname{endmcitethebibliography}  {\let\endmcitethebibliography\endthebibliography}{}

\end{document}